\documentclass[showpacs,10pt,twocolumn]{revtex4-1}

\usepackage{amsmath}
\usepackage{amssymb}
\usepackage{graphicx}
\usepackage{graphics}
\usepackage{epsfig}
\usepackage{CJK}
\usepackage{color}
\usepackage{epstopdf}
\usepackage{soul}

\usepackage{caption}

\begin{document}
\begin{CJK*}{GBK}{}

\title{Topological Quantum Materials with Kagome Lattice}
\author{Qi Wang$^{1,2,4}$, Hechang Lei$^{2,3,*}$, Yanpeng Qi$^{1,4,5,*}$, and Claudia Felser$^{6}$}
\affiliation{$^{1}$ School of Physical Science and Technology, ShanghaiTech University, Shanghai 201210, China\\
$^{2}$ Department of Physics and Beijing Key Laboratory of Opto-electronic Functional Materials $\&$ Micro-nano Devices, Renmin University of China, Beijing 100872, China\\
$^{3}$ Key Laboratory of Quantum State Construction and Manipulation (Ministry of Education), Renmin University of China, Beijing 100872, China\\
$^{4}$ ShanghaiTech Laboratory for Topological Physics, ShanghaiTech University, Shanghai 201210, China\\
$^{5}$ Shanghai Key Laboratory of High-Resolution Electron Microscopy, ShanghaiTech University, Shanghai 201210, China\\
$^{6}$ Max Planck Institute for Chemical Physics of Solids, Dresden 01187, Germany.
}

\date{\today}

\begin{abstract}
\textbf{CONSPECTUS:}
  
Recently, various topological states have undergone a spurt of progress in the field of condensed matter physics. An emerging category of topological quantum materials with kagome lattice has drawn enormous attention. A two-dimensional kagome lattice composed of corner-sharing triangles is a fascinating structural system, which could not only lead to geometrically frustrated magnetism, but also have a nontrivial topological electronic structure hosting Dirac points, van Hove singularities, and flat bands. 
Due to the existence of multiple spin, charge, and orbit degrees of freedom accompanied by the unique structure of the kagome lattice, the interplay between frustrated magnetism, nontrivial topology, and correlation effects is considered to result in abundant quantum states and provides a platform for researching the emergent electronic orders and their correlations.

In this account, we will give an overview of our research progress on novel quantum properties in topological quantum materials with kagome lattice. Here, there are mainly two categories of kagome materials: magnetic kagome materials and nonmagnetic ones. 
On one hand, magnetic kagome materials mainly focus on the 3$d$ transition-metal-based kagome systems, including Fe$_{3}$Sn$_{2}$, Co$_{3}$Sn$_{2}$S$_{2}$, YMn$_{6}$Sn$_{6}$, FeSn, and CoSn. The interplay between magnetism and topological bands manifests vital influence on the electronic response. 
For example, the existence of massive Dirac or Weyl fermions near the Fermi level significantly enhances the magnitude of Berry curvature in momentum space, leading to a large intrinsic anomalous Hall effect. 
In addition, the peculiar frustrated structure of kagome materials enables them to host a topologically protected skyrmion lattice or noncoplaner spin texture, yielding a topological Hall effect that arises from the real-space Berry phase. 
On the other hand, nonmagnetic kagome materials in the absence of long-range magnetic order include CsV$_{3}$Sb$_{5}$ with the coexistence of superconductivity, charge density wave state, and band topology and van der Waals semiconductor Pd$_{3}$P$_{2}$S$_{8}$.
For these two kagome materials, the tunability of electric response in terms of high pressure or carrier doping helps to reveal the interplay between electronic correlation effects and band topology and discover the novel emergent quantum phenomena in kagome materials.

\end{abstract}

\maketitle

\end{CJK*}

\textbf{1. INTRODUCTION}

\begin{figure*}
	\centerline{\includegraphics[scale=0.4]{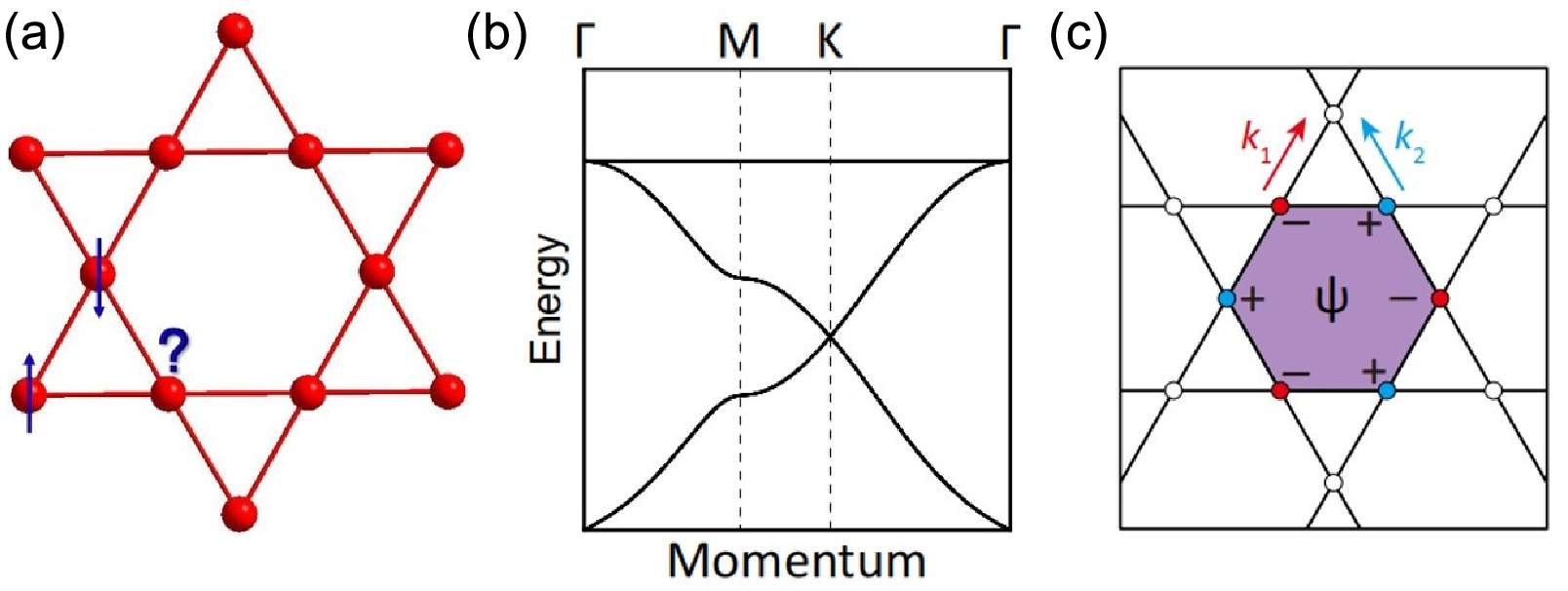}}
	\caption{(a) The geometrical structure of 2D kagome lattice. The blue arrows and question mark on a triangular sublattice denote the frustrated configuration. (b) Topological band structure of kagome lattice in consideration of the nearest electron hopping without SOC. (c) The schematic of destructive interference of Bloch wave functions ($\psi$) in kagome lattice. (b-c) Reproduced with permission from ref \cite{240}. Copyright 2021 Nature. }
\end{figure*}

Recently, topological quantum materials that host nontrivial topological states have become a hot frontier in condensed matter systems. Various intriguing topological states have been predicted theoretically and observed experimentally, including topological insulators, Dirac and Weyl semimetals, etc.\cite{1057,021004} Topological insulators are characterized by insulated bulk state with topologically protected conductive boundary/surface states. Dirac and Weyl semimetals belong to topological semimetals, which are a new class of topological quantum states featuring gapless band crossings of linear dispersion in the vicinity of the Fermi level. Dirac fermions that satisfy the Dirac equation are 4-fold degenerate and further break spatial-inversion or time-reversal symmetry, and 2-fold degenerate Weyl fermions appear in pairs and have opposite chirality. In topological quantum materials, owing to the intriguing nontrivial topological band structure, many fascinating properties have been observed, such as extremely large magnetoresistance, chiral anomaly and surface-state Fermi arcs, etc.

Topological quantum materials with kagome lattice have now become an emerging class of topological materials systems. A two-dimensional (2D) kagome lattice is geometrically defined as corner-sharing triangles, hosting a strong geometric frustration effect (Figure 1a). In the case of the antiferromagnetic Heisenberg model on a triangle sublattice, there is no unique ground state that can simultaneously satisfy the antiferromagnetic correlation interactions, leading to the emergence of spin disorder or strong quantum fluctuation. Due to the peculiar structural configuration, kagome lattice has been expected to be one of candidate systems to realize exotic quantum states, such as quantum spin liquid state. Moreover, with a decrease in the degree of degeneracy of the ground state, a noncollinear/coplaner spin frustrated configuration in kagome lattice is permitted to exist. 
In the view of electronic structure, considering the nearest electron hopping based on the tight-binding model in the absence of spin orbit coupling (SOC), the electronic band structure along high symmetry directions accordingly gives rise to three nontrivial energy bands (Figure 1b).\cite{240} Two of them are similar to the case in honeycomb lattice, where 2-fold degenerate bands cross linearly at the $K$ point in momentum space, generating a 4-fold degenerate Dirac point. 
The third band with zero bandwidth exhibits a dispersionless feature in the entire Brillouin zone (BZ), commonly referred to as a flat band. It is generated by the destructive interference of Bloch wave functions in the case of nearest electron hopping (Figure 1c), where the Bloch electrons localized within the hexagon of kagome lattice possess fully quenched kinetic energy, and Coulomb interactions dominant. The flat band and Dirac band touch and degenerate at the $\Gamma$ point. Furthermore, two van Hove singularities in the Dirac band are formed at the M point. 

The nontrivial topological characteristics and topological magnetic excitations are expected to play a vital role on the electronic response.
Generally, the measurement of the Hall effect has become a crucial means to characterize the electromagnetic response of conduction electrons in metallic materials. After Edwin H. Hall first discovered the normal Hall effect (NHE) in 1879 \cite{287}, which arises from the transverse motion of carriers caused by the Lorentz force, an additional large contribution of resistivity related to spontaneous magnetization was revealed in ferromagnetic metals two years later, known as the anomalous Hall effect (AHE).\cite{301,157} The Hall resistivity can be described by the empirical formula, 

\begin{equation}
\rho_{xy} = \rho_{xy}^{\rm N} + \rho_{xy}^{\rm A} = R_{0}B+R_{s}\mu_{0}M
\end{equation}

where the ordinary Hall coefficient $R_{0}$ determines the concentration and type of apparent carriers and $R_{s}$ corresponds to the anomalous Hall coefficient, which is usually much larger than $R_{0}$. 

Up to now, it is generally admitted that the intrinsic or extrinsic mechanism (side jump scattering and skew scattering) could account for the physical origin of AHE. On one hand, the intrinsic mechanism first proposed by Karplus and Luttinger is considered to be exclusively relevant to the inherent band structure of the material.\cite{1154} Later, one can understand the intrinsic AHE in the view of the Berry curvature effect.\cite{207208} The anomalous Hall conductivity (AHC) $\sigma_{xy}^{\rm A}$ can be determined by integrating the Berry curvature ($\Omega$) of all the occupied bands below the Fermi level ($E_{\rm F}$) : $\sigma_{xy}^{\rm A} = -\frac{2\pi e^{2}}{h}\int_{\rm BZ}\frac{d^{3}\textbf{k}}{(2\pi)^{3}}\sum_{n}f_{n}(\textbf{k})\Omega_{n}^{z}(\textbf{k})$, where $h$ and $f_{n}(\textbf{k})$ are the Plank constant and the Fermi-Dirac distribution function, respectively. Therefore, AHE should be observed in a system with time-reversal symmetry broken that hosts a nonzero Berry curvature. The intrinsic AHC manifests the maximum magnitude of $e^{2}/ha$ (10$^{2}$ - 10$^{3}$ $\Omega^{-1}$cm$^{-1}$) in the three-dimensional (3D) condition when $E_{\rm F}$ is within the energy gap, where $a$ is the lattice constant. In particularly, for kagome topological semimetals with broken time-reversal symmetry, the massive Dirac points with Chern gaps in the condition of Kane-Mele-type SOC or magnetic Weyl points will greatly enhance the magnitude of Berry curvature when the topologically protected linearly band crossings are close to $E_{\rm F}$, contributing to a large intrinsic AHE. On the other hand, extrinsic mechanisms are interpreted as arising from the impurity-induced asymmetrical spin scattering in the presence of spin-orbit interaction (SOI).\cite{39,4559} Generally, the extrinsic contribution is inescapable and usually coexists with intrinsic contribution in real materials. 

In addition to the effects of Berry curvature in momentum space, for noncoplanar spin texture such as topologically protected skyrmions state or noncoplanar magnetic structure with nonzero scalar spin chirality $\chi_{ijk}={\textit{\textbf{S}}}_{i}\cdot({\textit{\textbf{S}}}_{j}\times {\textit{\textbf{S}}}_{k})$ (where $\textit{\textbf{S}}_{i}$, $\textit{\textbf{S}}_{j}$, and  $\textit{\textbf{S}}_{k}$ are the three nearest spins), a real-space Berry phase can be obtained when the Bloch electrons cross the peculiar magnetic structure. Correspondingly, a fictitious magnetic field similar to the role of an extrinsic magnetic field in NHE emerges and generates a topological Hall effect (THE).\cite{156601} In magnetic kagome materials, the natural frustrated structure offers potentialities to realize magnetic excitations and THE.

Moreover, van Hove singularities and flat bands in kagome lattice both contribute to a large density of states (DOS), which could lead to strong electron correlation effects. This is in contrast with Dirac fermions, which make no contribution to DOS. Diverse correlated electronic states including superconductivity and ferromagnetism as well as Wigner crystal are predicted in the flat band model. Additionally, when the time-reversal symmetry is broken when considering SOC, a topological flat band analogous to the Landau level possesses a nonzero Chern number. It could give rise to a fractional quantum Hall effect at partial filling at high temperature (even room temperature).\cite{236802} With regard to van Hove singularities, the various electronic ground states like superconductivity or spin/charge density wave (CDW) can be realized at van Hove filling concomitant with varying Coulomb interaction.\cite{115135}

This account mainly focuses on the topological quantum materials that feature kagome lattice composed of diverse 3$d$/4$d$ transition metal atoms. According to our research results on the electrical transport responses combined with theoretical calculations and spectroscopy experiments, this Account will be described in four parts. 
In the first part, we provide an overview of the fundamental properties of kagome lattice as well as the correlated electromagnetic responses including AHE and THE. In the second and third parts, we will systematically introduce our research progress regarding the novel quantum properties in a series of representative topological kagome materials, consisting of magnetic kagome materials Fe$_{3}$Sn$_{2}$, Co$_{3}$Sn$_{2}$S$_{2}$, YMn$_{6}$Sn$_{6}$, FeSn, and CoSn as well as nonmagnetic kagome materials CsV$_{3}$Sb$_{5}$ and Pd$_{3}$P$_{2}$S$_{8}$. The last part will give a brief conclusion and perspective of topological kagome materials. 
 
\textbf{2. MAGNETIC KAGOME MATERIALS} 

\textbf{2.1 Fe$_{3}$Sn$_{2}$} 

\begin{figure*}
	\centerline{\includegraphics[scale=0.8]{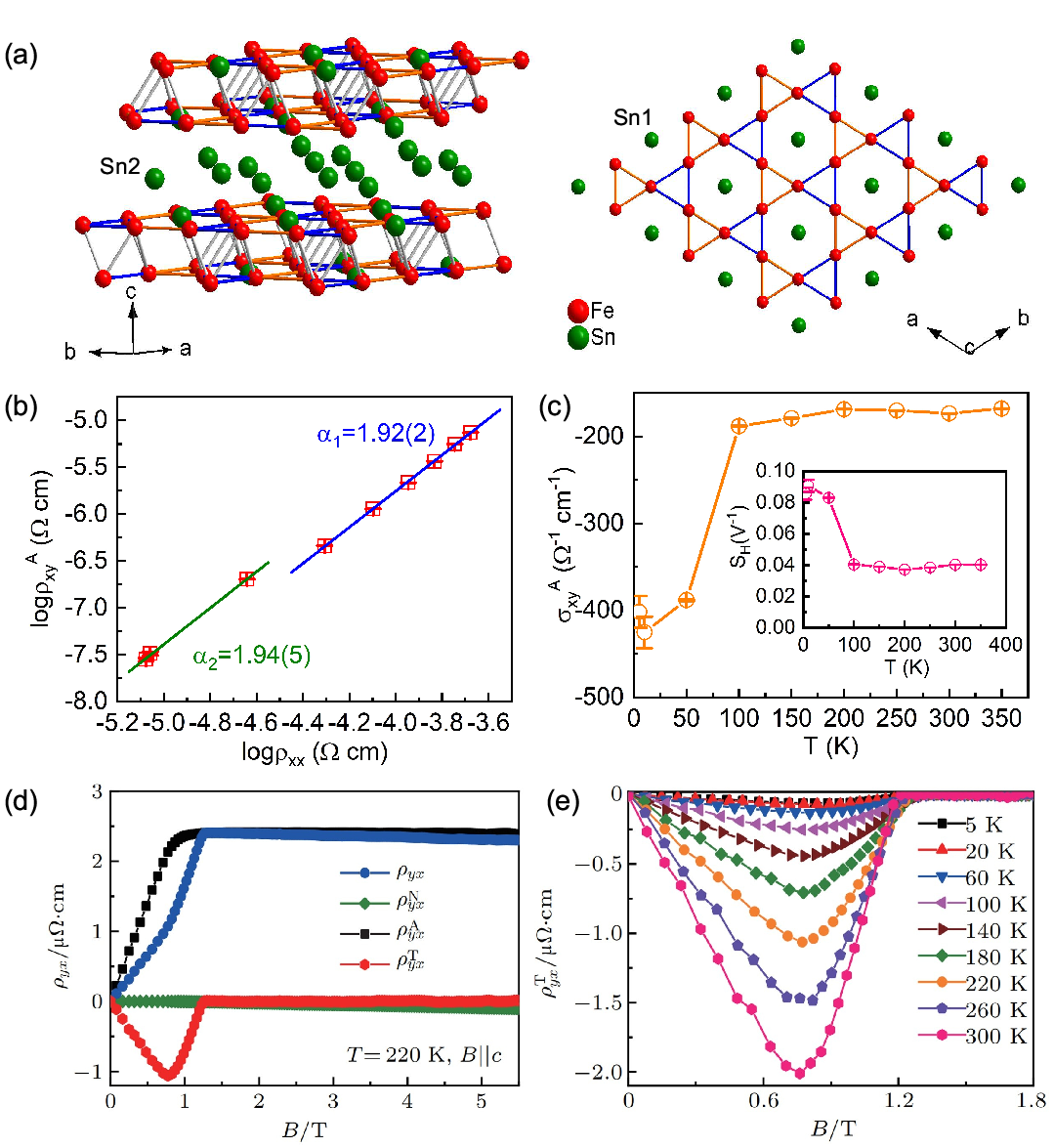}}
	\caption{(a) The crystal structure of Fe$_{3}$Sn$_{2}$. (b) Scaling relationship between $\rho_{xy}^{\rm A}$ and $\rho_{xx}$ in Fe$_{3}$Sn$_{2}$ single crystal. (c) Temperature dependence of $\sigma_{xy}^{\rm A}(T)$. The inset denotes $S_{\rm H}(T)$ curve. (d) Field-dependent $\rho_{yx}^{\rm N}(B)$, $\rho_{yx}^{\rm A}(B)$ and $\rho_{yx}^{\rm T}(B)$ at 220 K for $B\Vert c$. (e) $\rho_{yx}^{\rm T}(B)$ as a function of field between 5 and 300 K. Reproduced with permission from ref \cite{075135} and \cite{017101}. Copyright 2016 APS and 2020 IOP.}
\end{figure*} 

The first case of kagome magnets is the ferromagnet Fe$_{3}$Sn$_{2}$ with a Curie temperature $T_{\rm c}$ $\approx$ 640 K.\cite{075135} For the crystal structure, the bilayer kagome lattice composed of Fe atoms is sandwiched between Sn2 honeycomb layers along the $c$-axis (Figure 2a). Sn1 atoms are located in the centers of hexagons in each monolayer kagome lattice. The study of neutron powder diffraction revealed that there is a spin reorientation process in Fe$_{3}$Sn$_{2}$ where Fe spins tilt from $c$-axis to the $ab$ plane upon reducing temperature.\cite{452202} It has also been reflected in magnetic susceptibility curves, accompanying a plateau at low temperature under $H\Vert ab$, indicating the almost accomplishment of spin reorientation.\cite{075135} 

The Hall resistivity curves $\rho_{xy}(B)$ show the analogous shapes to those of $M(B)$ which display typical ferromagnetic characteristics, suggesting the presence of AHE. In accordance with formula (1), the negative $R_{0}$ yielded by the fitted slop of $\rho_{xy}(B)$ in the saturated field region demonstrates that the dominant carriers are electrons. 
As shown in Figure 2b, the near-quadratic scaling relation between $\rho_{xy}^{\rm A}$ and $\rho_{xx}$ demonstrates that the intrinsic or extrinsic side-jump mechanism instead of skew scattering ($\rho_{xy}^{\rm A} \propto \rho_{xx}$) is responsible for the AHE. 
Extraordinarily, it is found that $\sigma_{xy}^{\rm A}$ ($\approx-\rho_{xy}^{\rm A}/\rho_{xx}^{2}=-R_{s}\mu_{0}M/\rho_{xx}^{2}$) shows large value and approaches $-$400 $\Omega^{-1}$ cm$^{-1}$ when the temperature drops below 100 K (Figure 2c).
In theory, the contribution of side-jump scattering $|\sigma_{xy, \rm sj}^{\rm A}|$ could be estimated by means of $(e^{2}/ha)(\varepsilon_{\rm SO}/E_{\rm F})$, where $\varepsilon_{\rm SO}$ is spin-orbit interaction. 
The ratio of $\varepsilon_{\rm SO}$ to $E_{\rm F}$ is generally less than 10$^{-2}$, which would lead to a rather small AHE. 
Hence, the mainly contribution of AHE should arise from the intrinsic mechanism. Specially, the temperature-independent $S_{\rm H}$ ($\mu_{0}R_{s}/\rho_{xx}^{2}$ = $-\sigma_{xy}^{\rm A}/M$) also confirmed the intrinsic origin (inset of Figure 2c). 
Furthermore, in 2018, the study of angle-resolved photoemission spectroscopy (ARPES) by Checkelsky's group revealed that there are massive Dirac fermions with a gap of 30 meV near $E_{\rm F}$ in Fe$_{3}$Sn$_{2}$ due to SOC and the Berry curvature is enormously enhanced, contributing to a large AHC.\cite{638} 

Generally, magnetic skyrmions induced by Dzyaloshinskii-Moriya interaction are commonly observed in noncentrosymmetric systems, like B20 chiral magnets.\cite{186602}
Recently, the existence of skyrmions has been extended into centrosymmetric magnetic materials, especially for the frustrated magnets where skyrmions are stabilized by magnetic frustration interaction.\cite{6456} Interestingly, Lorentz transmission electron microscopy (LTEM) reported the presence of field-induced skyrmionic bubbles in centrosymmetric frustrated Fe$_{3}$Sn$_{2}$ over a wide temperature range from 100 K to 400 K.\cite{1701144} Moreover, it is noticed that the slopes between $\rho_{yx}(B)$ and $M(B)$ below the saturated field exhibit a slight difference, suggesting the possible additional contribution of THE to Hall resistivity.\cite{017101} Considering the topological Hall component, the Hall resistivity can be further expressed in three parts: 

\begin{equation}
\rho_{\rm H} = \rho_{\rm H}^{\rm N} + \rho_{\rm H}^{\rm A} + \rho_{\rm H}^{\rm T} = R_{0}B + S_{\rm H}\rho^{2}M + \rho_{\rm H}^{\rm T}
\end{equation}

where $\rho_{\rm H}^{\rm T}$ is the topological Hall resistivity. From the analyses of scaling curves, the component of THE is undoubtedly obtained below 1.3 T in a wild temperature range (5 $-$ 300 K) as shown in Figure 2d and 2e. The monotonically varying $\rho_{yx}^{\rm T}$ possesses a giant topological Hall response with a maximum absolute value of 2.01 $\mu\Omega$ cm at 300 K. 
The magnitude is comparable with that in centrosymmetric Gd$_{2}$PdSi$_{3}$ (2.6 $\mu\Omega$ cm at 2 K) with skyrmion lattice \cite{6456}, significantly larger than those in previous reported skyrmionic B20 materials like MnSi (4 n$\Omega$ cm).\cite{186602} 
The estimated helical period in view of skyrmionic bubbles is found to be 8.8 nm, which is considerably smaller than that ($\sim$ 500 nm) observed in LTEM. Considering the presence of spin reorientation that could generate noncollinear magnetic structure, it can be concluded that the synergistic effect of noncollinear spin texture and skyrmion structure is possibly responsible for the giant THE. On the other hand, above the field of 1.3 T, THE disappears own to the full polarization of Fe spins toward the field, then intrinsic AHE dominants in the Hall response (Figure 2d).

Furthermore, a neutron scattering study revealed that the in-plane local magnetism in Fe$_{3}$Sn$_{2}$ can be efficiently regulated by small magnetic fields, generating a pronounced electromagnetic response that shows anisotropic magnetoresistance (MR) under varying in-plane magnetic fields.\cite{196604} Additionally, Zeng, $et$ $al$. confirmed the existence of flat band below $E_{\rm F}$ about 0.2 eV by utilizing ARPES.\cite{096401} 

\textbf{2.2 Co$_{3}$Sn$_{2}$S$_{2}$} 

\begin{figure*}
	\centerline{\includegraphics[scale=0.3]{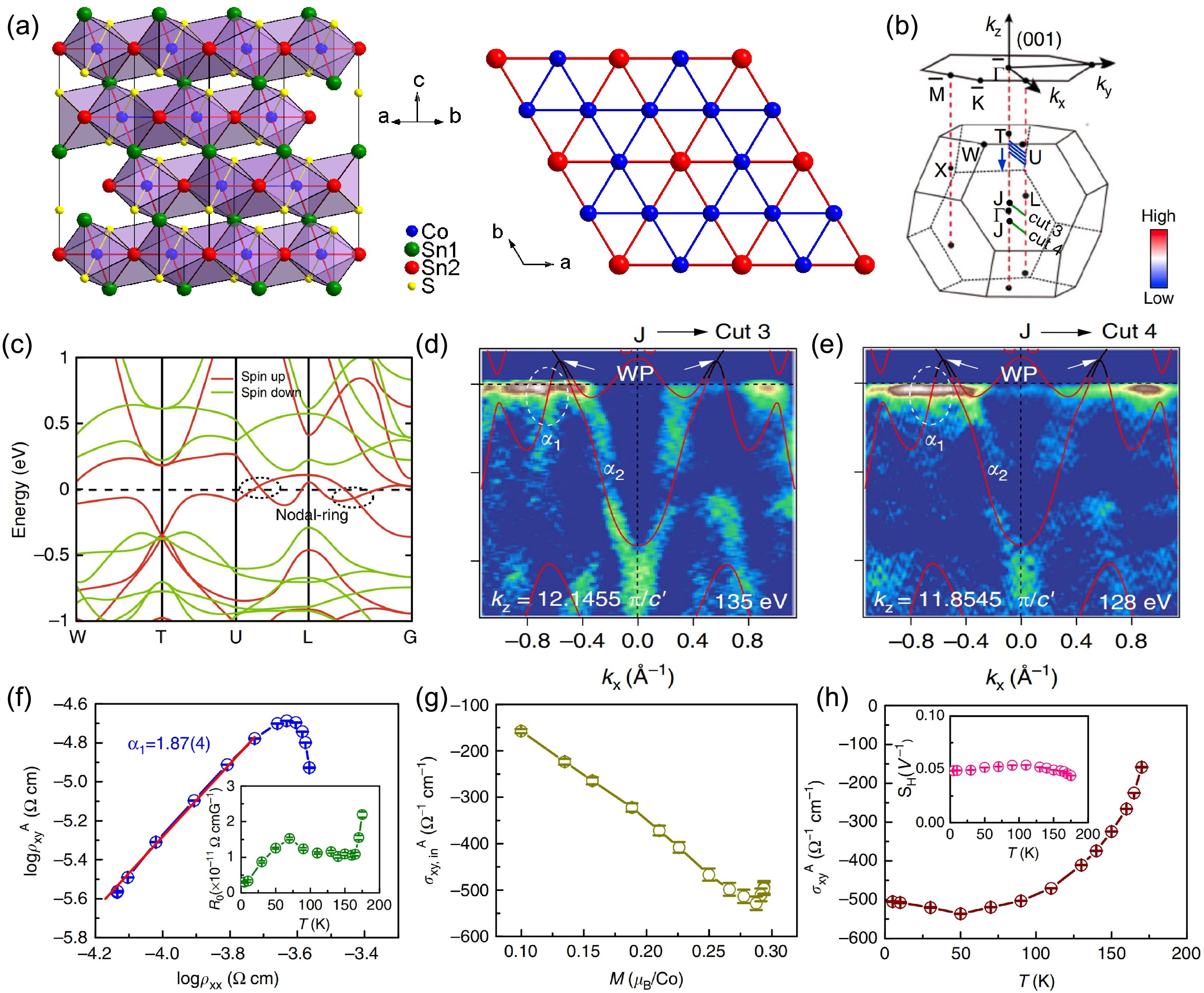}}
	\caption{(a) Crystal structure of Co$_{3}$Sn$_{2}$S$_{2}$. (b) The schematic of 3D and 2D projected BZ. (c) The calculated band structure without spin orbit coupling. (d and e) ARPES spectra along cuts 3 and 4 of the Co$_{3}$Sn$_{2}$S$_{2}$ single crystal and the corresponding theoretical calculations (red lines). (f) log$\rho_{xy}^{\rm A}(T)$ versus log$\rho_{xx}(T)$; the inset is the $R_{0}(T)$ curve. (g) $M$ dependence of $\sigma_{xy,\rm{in}}^{\rm A}$. (h) Temperature dependence of $\sigma_{xy}^{\rm A}$. Reproduced with permission from ref \cite{3681}. Copyright 2018 Nature. }
\end{figure*}

Various kinds of topological semimetals have been discovered by means of experiments and theories, while the studies on magnetic Weyl semimetals with breaking time-reversal symmetry make relatively slow progress. 
HgCr$_{2}$Se$_{4}$ and Y$_{2}$Ir$_{2}$O$_{7}$ were predicted to host magnetic Weyl fermions in the early theoretical calculations \cite{205101,186806}, but have not yet been confirmed experimentally. Due to the existence of magnetic interaction and the Weyl fermions being generally located away from the $E_{\rm F}$ in real materials, the realizations of magnetic Weyl fermions and the corresponding electromagnetic response are challenging. Consequently, the search for magnetic Weyl fermions has remained stagnant. The breakthrough in magnetic Weyl semimetals stems from the discovery of half-metallic ferromagnet Co$_{3}$Sn$_{2}$S$_{2}$, found independently by Lei's group and Felser's group.\cite{3681,1125}

Co$_{3}$Sn$_{2}$S$_{2}$ features a Co kagome layer in which Sn2 atoms are situated in the centers of hexagons (Figure 3a). It experiences a ferromagnetic transition at $T_{\rm c}$ $\approx$ 174 K and displays a soft ferromagnetic behavior characterized by a small coercive field of 0.08 - 0.1 T. Due to the half-metallic feature, Co$_{3}$Sn$_{2}$S$_{2}$ manifests a full polarization of spins toward one direction. 
The 3$d$ orbitals of Co atom mainly contribute to the bands close to $E_{\rm F}$, which form a nodal ring around the $L$ point in the absence of SOC (Figure 3c). When introducing SOC, three pairs of Weyl points with opposite chirality located at 50 meV above $E_{\rm F}$ and away from the high-symmetry directions are obtained.
On the other hand, ARPES experiments were performed to characterize the electronic structure in Co$_{3}$Sn$_{2}$S$_{2}$. There is a trend of linear dispersion between the $\alpha_1$ and $\alpha_2$ bands above $E_{\rm F}$ along the cuts 3 and 4 (Figure 3d and 3e). The consistence between the ARPES results and theoretical calculations confirms the existence of Weyl fermions in Co$_{3}$Sn$_{2}$S$_{2}$.  

It is reported that the Weyl fermions in magnetic Weyl semimetal acting as magnetic monopoles in momentum space can make vital contributions to Berry curvature. 
It can lead to the remarkable transport response like huge AHE, and thus we carried out the transport measurements on single crystal Co$_{3}$Sn$_{2}$S$_{2}$. The hole-type carriers are dominant in Co$_{3}$Sn$_{2}$S$_{2}$ due to the positive $R_{0}$ (inset of Figure 3f). As shown in Figure 3f, $\rho_{xy}^{\rm A}$ almost scales quadratically with $\rho_{xx}$ at low temperatures, which firstly excludes the contribution of skew scattering. 
On the other hand, the intrinsic contribution under various temperatures can be extracted by the scaling curves of $\rho_{xy}^{\rm A}/(\rho_{xx}M)$ versus $\rho_{xx}$, according to the formula $\rho_{xy}^{\rm A}=a(M)\rho_{xx}+b(M)\rho_{xx}^{2}$. Consequently, a remarkable large intrinsic AHC $|\sigma_{xy,\rm{in}}^{\rm A}|$ was obtained, reaching a value as much as 505 $\Omega^{-1}$ cm$^{-1}$ at low temperature (Figure 3g). Moreover, the $\sigma_{xy,\rm in}^{\rm A}$ depends linearly on magnetization, which is well consistent with the results of theoretical calculations. By contrast, the estimated $|\sigma_{xy, \rm{sj}}^{\rm A}|$ only contributes a small magnitude of 3.9 $\Omega^{-1}$ cm$^{-1}$.
Another intrinsic feature is the temperature-independent behavior of $\sigma_{xy}^{\rm A}$ and $S_{\rm H}$ (Figure 3h). 
Moreover, the $|\sigma_{xy}^{\rm A}|$ decreases gradually when $T >$ 140 K, which primarily arises from the obvious reduction in $M(T)$ near $T_{\rm c}$. The scaling analyses confirm that the striking AHE is dominated by an intrinsic mechanism. 
Combined with the first-principles calculations and ARPES, the large intrinsic AHC emerged in Co$_{3}$Sn$_{2}$S$_{2}$ is closely related to the presence of magnetic Weyl fermions close to $E_{\rm F}$. Later on, the observations of Fermi arcs and related phenomena by using APRES and scanning tunneling microscope (STM) in other research groups further confirm the existence of Weyl fermions in ferromagnetic Co$_{3}$Sn$_{2}$S$_{2}$.\cite{1282, 1278} In addition, a giant anomalous Hall angle (20 $\%$) and negative MR associated with chiral anomaly of Weyl fermions have also been observed.\cite{1125} 

Regarding the physics of flat band in Co$_{3}$Sn$_{2}$S$_{2}$, STM measurements demonstrate a sharp peak in both the d$I$/d$V$ spectra on S and side surface below $E_{\rm F}$, which arises from the contribution of the flat band in kagome lattice.\cite{443} The peak shows a positive energy shift under magnetic fields which is distinct from the conventional Zeeman effect, ascribing to the negative orbital magnetism of the flat band. These experimental studies deepen the understanding of magnetic Weyl semimetals significantly.

\textbf{2.3 YMn$_{6}$Sn$_{6}$} 

\begin{figure*}
	\centerline{\includegraphics[scale=0.70]{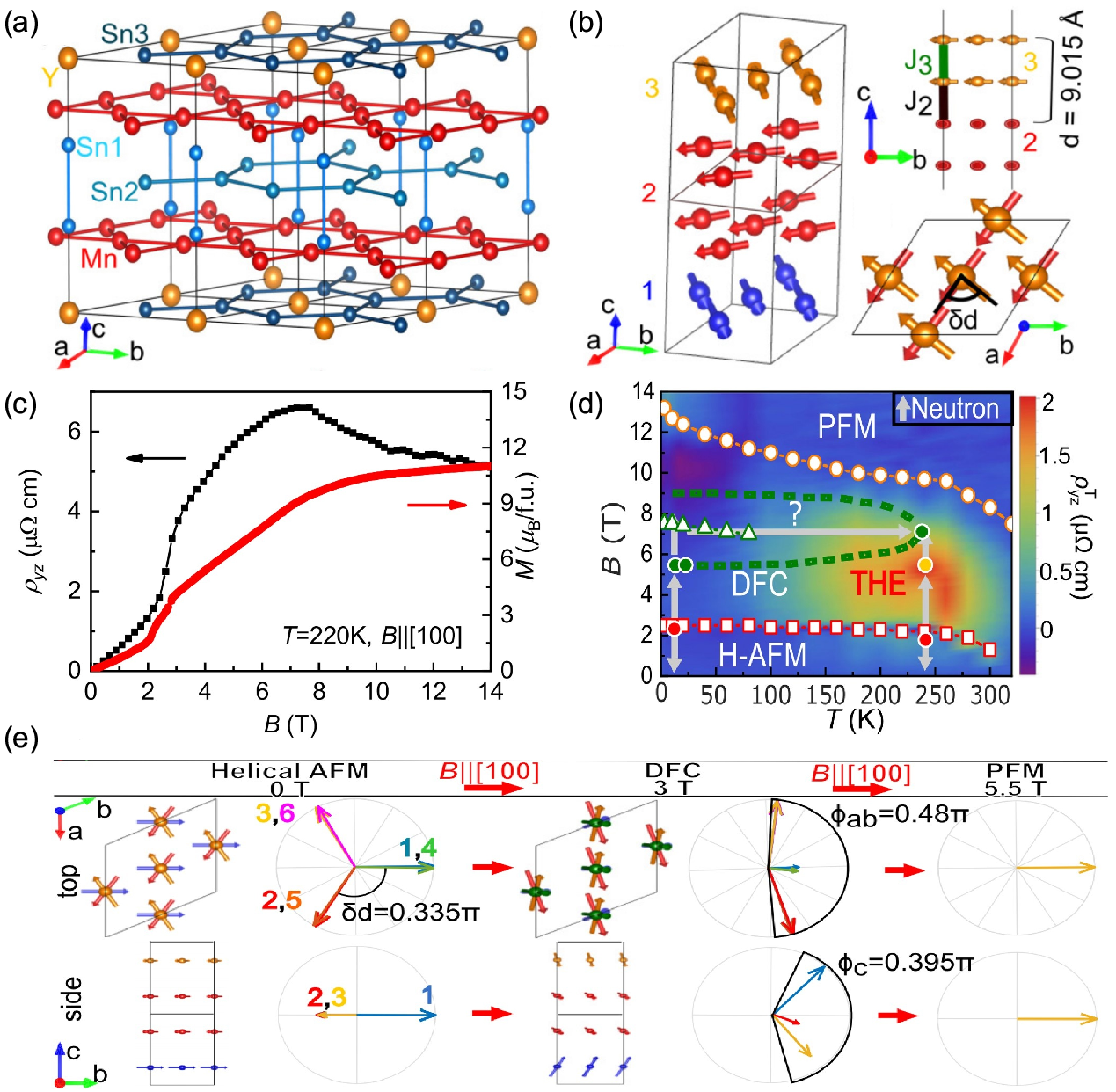}}
	\caption{(a and b) Crystal structure (left) and zero-field magnetic structure (right) of YMn$_{6}$Sn$_{6}$. $d$ displays the distance between the nearest kagome bilayer. $J_{2}$ and $J_{3}$ represent the magnetic exchange couplings in adjacent layers. (c) Field-dependent Hall resistivity $\rho_{yz}(B)$ and $M(B)$ in the field of the $[100]$ direction. (d) Phase diagram along the $[100]$ direction field. YMn$_{6}$Sn$_{6}$ undergoes a helical antiferromagnetic (H-AFM) structure, a double-fan structure with a $c$-axis component (DFC), and a polarized ferromagnetic (PFM) structure under magnetic field. (e) Magnetic structure model of YMn$_{6}$Sn$_{6}$ at a temperature of 240 K. Reproduced with permission from ref \cite{014416}. Copyright 2021 APS.}
\end{figure*}

The centrosymmetric antiferromagnet YMn$_{6}$Sn$_{6}$ belongs to RMn$_{6}$Sn$_{6}$ (R = rare earth elements) kagome family with space group $P6/mmm$. In contrast to the crystal structures of the kagome magnets Fe$_{3}$Sn$_{2}$ and Co$_{3}$Sn$_{2}$S$_{2}$ where Sn atoms are located in the kagome layer, the occupation of R atoms in R-Sn3 layer leads to the displacement of Sn1 atoms out of the Mn kagome layer, resulting in the emergence of a clean Mn kagome lattice (Figure 4a).\cite{014416} The family exhibits multiple magnetic ground states by varying R elements, like ferrimagnets RMn$_{6}$Sn$_{6}$ (R = Gd - Ho), antiferromagnets ErMn$_{6}$Sn$_{6}$ and YMn$_{6}$Sn$_{6}$.\cite{014416,246602}
In YMn$_{6}$Sn$_{6}$, the magnetization only arises from the contribution of Mn atoms. The Mn spins in each kagome bilayer (Mn-Sn1-Sn2-Sn1-Mn sheets) are in a linear ferromagnetic arrangement without $c$-axis component and collectively rotate $\delta d$ degree to the adjacent kagome bilayer (Figure 4b), i.e., YMn$_{6}$Sn$_{6}$ exhibits a helical antiferromagnetism ($T_{\rm N}\approx$ 333 K). When applying field along the $ab$-plane, a spin-flop transition appears around 2.5 T below 300 K. 

From the measurements of Hall resistivity, it is found that the carrier densities of electrons and holes change when applying fields along $ab$-plane and $c$-axis, reflecting the obvious influence of field on the electrical structure. The relatively large intrinsic AHE was realized in YMn$_{6}$Sn$_{6}$ for both field directions based on the scaling analyses between $\rho_{\rm H}/B$ and $\rho^{2}M/B$ and the weak dependence of $\sigma_{\rm H}^{\rm A}$ on longitudinal conductivity, featuring 45 S cm$^{-1}$ for $\sigma_{xy}^{\rm A}$ and 300 S cm$^{-1}$ for $\sigma_{zy}^{\rm A}$. These values are comparable with that in ferrimagnetism TbMn$_{6}$Sn$_{6}$ (121 $\pm$ 6 $\Omega^{-1}$ cm$^{-1}$).\cite{533} The Chern-gapped Dirac fermions realized by ARPES experiments,\cite{3129} as also observed in TbMn$_{6}$Sn$_{6}$ via STM, contribute to the large intrinsic AHC.  

Moreover, it is noticed that an obvious difference between $\rho_{\rm H}(B)$ and $M(B)$ exists below the fully spin polarized region. The signals of THE were obtained in different field directions after subtracting the NHE and AHE components. Surprisingly, the THE is larger when the field is parallel to the kagome plane, reaching a maximum $\rho_{yz}^{\rm T}$ of 2.0 $\mu\Omega$ cm at 5 T and 240 K. Whereas $\rho_{yx}^{\rm T}$ is relatively small with a maximum value $-$0.28 $\mu\Omega$ cm for the $c$-axial field of 5 T at 220 K. 
For the microscopic mechanism of the field-direction-dependent large THE, small-angle neutron scattering experiments first exclude the existence of skyrmion lattice in YMn$_{6}$Sn$_{6}$. A phase diagram, which integrates the magnetic structures determined by neutron diffraction and the topological Hall response $\rho_{yz}^{\rm T}$, is depicted in Figure 4d. For the $ab$-plane magnetic field, the Mn spins in the helical antiferromagnetic (H-AFM) structure are initially located in the kagome plane, yielding a small contribution to THE. 
Above the spin-flop transition field, a field-driven double-fan phase with a $c$-axis component (DFC) emerges (Figure 4e), generating a large nonzero spin chirality $\chi$ which is essential to the significant THE. 
The specific spin texture manifests exclusively in the case of the comparability of $c$-axis AFM exchange couplings and in-plane interactions. 
THE finally disappears in the fully polarized ferromagnetic (PFM) state at high filed, and then the intrinsic AHE dominates in YMn$_{6}$Sn$_{6}$. By contrast, the small $\rho_{yx}^{\rm T}$ for $H\Vert c$ could be related to the noncollinear magneic structure formed during the process that causes the in-plane Mn spin to progressively align toward the $c$-axis direction. 
In addition, for the characterization of the electronic structure, the flat band and saddle point near $E_{\rm F}$ were confirmed by ARPES.\cite{3129}

Another representative Mn-based kagome magnet is the antiferromagnetic Weyl semimetal Mn$_{3}$Sn, which shows a noncollinear magnetic order with in-plane 120$^{{\circ}}$ Mn spins. Nakatsuji's group reported a large intrinsic AHE by electrical transport experiments.\cite{212} In addition, the chiral anomaly associated with Weyl fermions was also observed.\cite{1090}

\textbf{2.4 FeSn and CoSn} 

FeSn and CoSn feature an isostructural crystal structure with space group $P6/mmm$, which is equivalent to turning the bilayer kagome lattice in Fe$_{3}$Sn$_{2}$ into a single Fe/Co kagome layer. However, there are different magnetic ground states in them. FeSn exhibits an antiferromagnetism with the $c$-axial antiferromagnetic and $ab$-plane ferromagnetic spin configurations. But for CoSn, it shows a Pauli paramagnetism. No anomalous electromagnetic responses emerged in both FeSn and CoSn. Interestingly, spectroscopic measurements revealed exotic quantum states and correlation effects.\cite{163,4002,4003} The coexistence of Dirac fermions and flat bands has been discovered via ARPES experiments in FeSn and CoSn.\cite{163,4002} In CoSn, the STM measurements demonstrated that the fermion-boson many-body interaction is driven by electrons coupling with the phonon flat band in the Co kagome lattice.\cite{4003} The evolution of spin excitations was revealed in FeSn and CoSn via inelastic neutron scattering experiments.\cite{240}

\textbf{3. NONMAGNETIC KAGOME MATERIALS} 

\textbf{3.1 CsV$_{3}$Sb$_{5}$} 

\begin{figure*}
	\centerline{\includegraphics[scale=0.65]{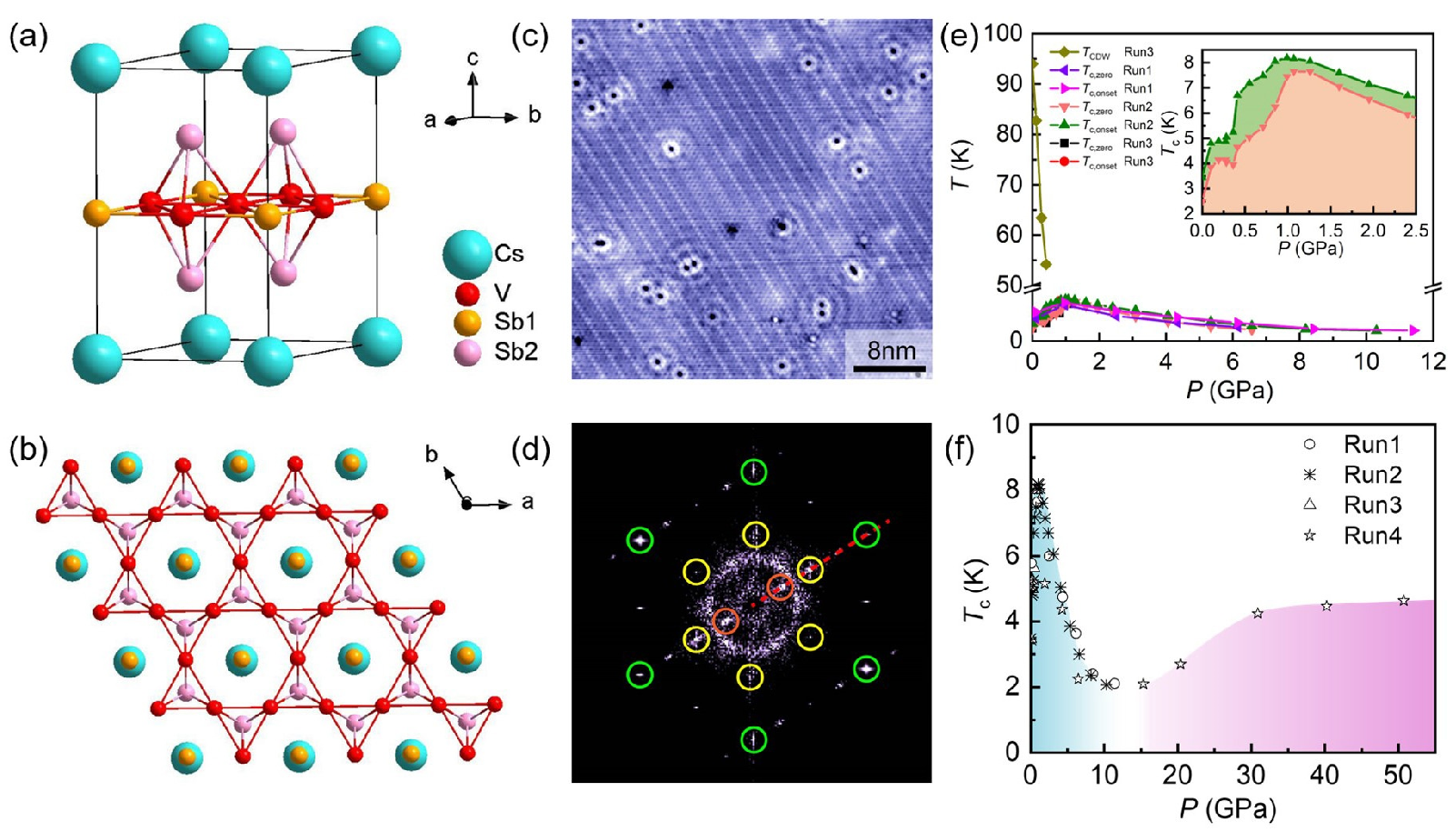}}
	\caption{(a and b) Crystal structure of CsV$_{3}$Sb$_{5}$. (c) High-resolution STM topography of Sb-terminated surface. (d) The corresponding Fourier transform. The green, yellow, and orange circle symbols denote Q Bragg peaks, Q$_{2a}$ and Q$_{4a}$ CDW vectors, respectively. (e and f) Phase diagrams of temperature ($T_{\rm c}$ and $T_{\rm CDW}$) and pressure ($P$) in CsV$_{3}$Sb$_{5}$. Reproduced with permission from ref \cite{2102813}. Copyright 2021 Wiley.}
\end{figure*}

Superconductivity is a long-term research hotspot in condensed matter physics. Especially when a system simultaneously hosts superconducting and topological characteristics, more exotic quantum effects could emerge, such as topological superconductivity, Majorana zero modes, etc. Although theoretical predictions suggest the existence of superconductivity in kagome system, the experimental studies on the superconductivity are still lacking. Until recently, the discoveries of the AV$_{3}$Sb$_{5}$ (A = K, Rb, Cs) family containing V kagome lattice have greatly advanced the study of superconductivity in kagome materials.\cite{094407,247002,037403} The crystal structure of AV$_{3}$Sb$_{5}$ consists of alternating A-Sb2-VSb1-Sb2-A layers, as shown in Figure 5a and 5b.
These compounds, as distinct from the 3$d$ transition-metal-based kagome magnets, have no magnetic order. Especially, they are rare kagome materials that simultaneously possess superconductivity ($T_{\rm c}$ $\approx$ 0.92 - 2.5 K) and CDW state ($T_{\rm CDW}$ $\approx$ 80 - 110 K) as well as a $Z_{2}$ topological band structure. Subsequently, a large number of experimental and theoretical works were carried out on AV$_{3}$Sb$_{5}$.  

Taking CsV$_{3}$Sb$_{5}$ as an example\cite{2102813}, a 2\textbf{a}$\times$2\textbf{a} CDW state on the Sb-terminated surface was identified in our STM measurements (yellow circles in Figure 5d). 
It is further revealed by other groups that the superlattice also has a $c$-axis component, i.e, a 3D CDW with a 2\textbf{a}$\times$2\textbf{a}$\times$2\textbf{c} pattern, according to STM, X-ray scattering and nuclear magnetic resonance experiments.\cite{031026,031050,247462} The emergence of CDW is closely related to the van Hove singularities at M points near E$_{F}$. Particularly, a strip-like unidirectional superstructure with a 4\textbf{a}$\times$1\textbf{a} vector has also been distinguished (orange circles in Figure 5d). The two types of CDW orders were both revealed in RbV$_{3}$Sb$_{5}$, while the latter 4\textbf{a}$\times$1\textbf{a} superlattice was absent in KV$_{3}$Sb$_{5}$. 
Recently, various CDW states were also observed in kagome antiferromagnet FeGe and paramagnet ScV$_{6}$Sn$_{6}$.\cite{490,216402} These studies provide meaningful guidance for exploring and understanding the CDW and nontrivial band topology in kagome systems.

It is generally acknowledged that high pressure has served as a clean and effective technique to tune in situ the crystal and electronic structure, with the expectation of realizing novel quantum phenomena.      
The peculiar AV$_{3}$Sb$_{5}$ compounds provide an ideal platform to investigate the evolution and relationship between superconductivity and CDW by utilizing high pressure. 
One can see that the electrical responses are very evident under the application of pressure (Figure 5e and 5f). 
As shown in the $T$-$P$ phase diagrams, $T_{\rm CDW}$ shows a rapid and monotonic decrease, while the superconducting transition temperature $T_{\rm c}$ exhibits superconducting domes under the pressure up to 50.7 GPa and attains the maximum value (8.2 K) at $\sim$ 1 GPa. With increasing pressure, superconductivity first disappears at 13.1 GPa and surprisingly re-enters a superconducting state above 15 GPa. In the whole pressure region, no structural phase transition is revealed. The enhancement of superconductivity at low pressures arises from the strongly competitive relation with CDW state. Upon compression, the decrease of displacements of Sb2 atoms is possibly responsible for the suppression of CDW order. 
Moreover, the emergence of the second superconducting state is closely associated with the changes in electronic structure. The dome-shaped superconducting behaviors have also emerged in RbV$_{3}$Sb$_{5}$ and KV$_{3}$Sb$_{5}$. 

Due to the weak interlayer interaction, the layered kagome structure is easier to mechanically exfoliate. It is found that $T_{\rm c}$ shows a nonmonotonic dependence on the thickness in CsV$_{3}$Sb$_{5}$.\cite{105} Nonreciprocal charge transport behaviors were observed in CsV$_{3}$Sb$_{5}$ thin flakes, exhibiting pronounced second harmonic signals in the superconducting state. It may be related to the intrinsic symmetry breaking arising from the unconventional superconducting state.

Although in the absence of magnetic order in AV$_{3}$Sb$_{5}$, Ali's group reported that there is a large AHC in KV$_{3}$Sb$_{5}$.\cite{eabb6003} The chiral CDW breaking the time-reversal symmetry is possibly associated with the emergence of AHE.\cite{1038} In addition, more intriguing correlated phenomena have been successively reported by other groups, such as pair density wave, nematic phase, quantum transport, 2-fold symmetry of resistivity, etc.\cite{222,59,207002,6727}

\textbf{3.2 Pd$_{3}$P$_{2}$S$_{8}$}

\begin{figure*}
	\centerline{\includegraphics[scale=0.40]{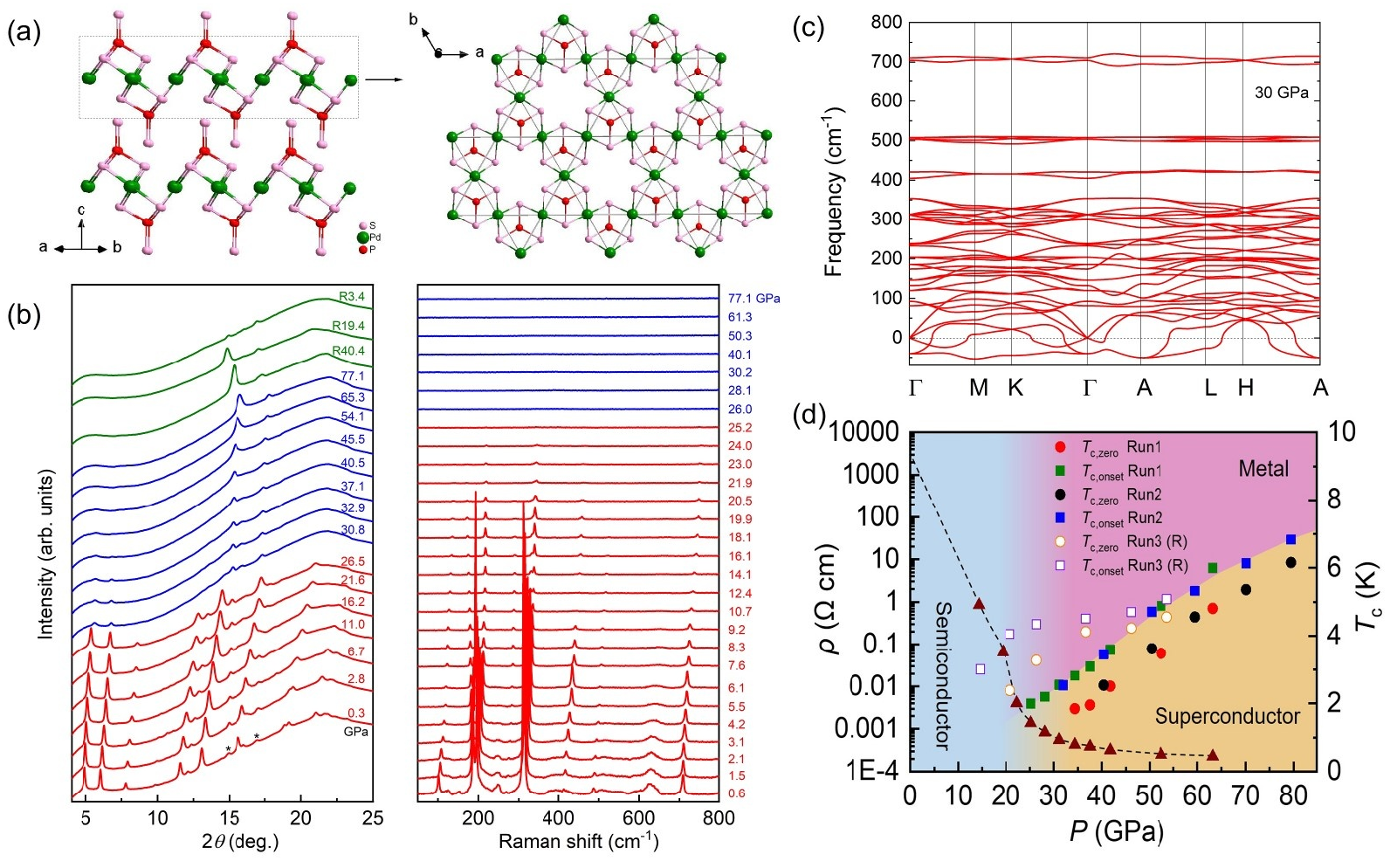}}
	\caption{(a) Crystal structure of Pd$_{3}$P$_{2}$S$_{8}$. (b) High-pressure synchrotron radiation XRD patterns (left) and Raman spectra (right). The asterisks display the diffraction peaks of rhenium gasket which arise from the trailing of the x-ray spot. R denotes the process of releasing pressure. (c) Phonon spectrum of Pd$_{3}$P$_{2}$S$_{8}$ at 30 GPa. (d) Phase diagram of Pd$_{3}$P$_{2}$S$_{8}$ under pressure. The values of resistivity $\rho$ are taken at 10 K. Reproduced with permission from ref \cite{043001}. Copyright 2023 IOP. }
\end{figure*}

Another kind of nonmagnetic kagome material is the van der Waals (vdW) compound Pd$_{3}$P$_{2}$S$_{8}$, which features a pristine Pd kagome lattice without other atoms occupations (Figure 6a).\cite{043001} Similarly to AV$_{3}$Sb$_{5}$, Pd$_{3}$P$_{2}$S$_{8}$ with weak interlayer vdW interaction can be easily cleaved into thin layers, even monolayer.\cite{20998} Nevertheless, Pd$_{3}$P$_{2}$S$_{8}$ is a semiconductor possessed an energy gap of 1.4 - 1.8 eV at ground state.\cite{20998,155115} 

Due to the intrinsic semiconducting characteristics, it is expected to tune the physical properties by means of high pressure or chemical doping. 
High-pressure electrical transport studies revealed that the semiconducting behavior can be gradually suppressed, ultimately transitioning into a metallic state at 34.4 GPa (Figure 6d). Simultaneously, the superconductivity is successfully induced in the pressurized Pd$_{3}$P$_{2}$S$_{8}$. The $T_{\rm c}$ increases monotonously with pressure, and a maximum and unsaturated $T_{\rm c}$ of 6.83 K is obtained within the limit of our research.
The almost vanishing strength of peaks above 26 GPa in both synchrotron radiation XRD and Raman spectroscopy suggest the occurrence of an amorphous phase transition when Pd$_{3}$P$_{2}$S$_{8}$ enters into superconducting state (Figure 6b). 
Accordingly, the emergence of pressure-induced superconductivity is closely related to the amorphous phase. The calculated phonon spectra of Pd$_{3}$P$_{2}$S$_{8}$ reveal the presence of imaginary modes above 20 GPa (Figure 6c), confirming the structural instability under high pressure in the superconducting state. 
Upon compression, the sharp decrease of the distance between interlayer Pd and S atoms significantly enhances the interlayer coupling, the crystal structure can be easily tuned into 3D and may become unstable. On the other hand, with Se doping at the S sites, the semiconducting ground state remains robust despite a gradual decrease of the band gap caused by the enhanced interlayer coupling.\cite{155115} 

\textbf{4. SUMMARY AND OUTLOOK} 

Kagome lattice is a fantastic geometrical structure, which contains a various of intriguing physical characteristics. It provides an excellent platform to study the exotic magnetism, quantum states, nontrivial topological properties and correlation effects. Here, we have reviewed the progress on the representative topological quantum materials with magnetic/nonmagnetic kagome lattice. From the perspective of transport behaviors of conduction electrons, combined with the theoretical calculations and spectroscopic measurements, various striking quantum phenomena have been revealed, such as momentum/real-space Berry phase-driven giant electromagnetic responses including AHE and THE, enhanced and even induced superconductivity under pressure, CDW states, negative flat band magnetism, and tunable local magnetism.
Our works undoubtedly enrich the variety of kagome material family and provide insight into the understanding of strong entanglement between frustration, topology, and correlation.

Benefit from the unique layered structure of kagome materials, it also provides a promising platform to explore low-dimensional physical performances. 
Although the most recently investigated kagome magnets tend to exhibit 3D bulk characteristics owing to the strong interlayer interactions, the nonmagnetic kagome materials like AV$_{3}$Sb$_{5}$ and Pd$_{3}$P$_{2}$S$_{8}$ that can be easily exfoliated down to a few layers or a single layer.
In the 2D limit, multiple experimental techniques, such as mechanical exfoliation, ionic liquid gating, intercalation, etc, can be utilized to manipulate the quantum states and discover novel physical phenomena. For example, by modulating the thickness of the film, it is expected to investigate the thickness-dependent AHE in kagome materials, including the varying magnitude of AHE and the evolution of the dominant mechanism. The dimensionality may also produce other exotic responses of kagome systems, such as the enhancement of superconductivity. Furthermore, the physical properties could also be modified through ionic liquid gating or intercalation, accompanied by the tuning of carrier densities, as well as the size and shape of the Fermi surface.

Theoretical calculations proposed that 2D kagome lattice with breaking time-reversal symmetry caused by magnetic order can realize a quantum anomalous Hall effect (QAHE), when the electron filling is within the gap. Magnetic Weyl semimetals including Co$_{3}$Sn$_{2}$S$_{2}$, Co$_{3}$Pb$_{2}$S$_{2}$, Co$_{3}$Pb$_{2}$Se$_{2}$ and Co$_{3}$Sn$_{2}$Se$_{2}$ were predicted to manifest nonzero Chern numbers and chiral edge states in the 2D limit, where Weyl points are broken and an energy gap opens.\cite{115106,014410} However, due to the difficulties in experimentally achieving an isolated magnetic 2D kagome layer, the QAHE accessing quantized AHC has not yet been observed in real kagome materials. In the future, the discovery of the suitable 2D magnetic kagome materials holds the promise of realizing this scenario. Moreover, the achievement of the quantized Hall effect enables to possess a large anomalous Hall angle $\sim$ 90$^\circ$.

The effects of topological Dirac/Weyl fermions on the transport properties have been intensively studied. Currently, the related investigations of natural flat band in the aspect of electrical properties are still sparse. Especially, the theoretically predicted superconductivity associated with flat band has not been realized in kagome materials, although flat bands near $E_{\rm F}$ as revealed in FeSn and CoSn. 
The appropriate position of flat band close to $E_{\rm F}$ with certain electron-electron correlations may be vitally important for the presence of superconducting state, and further studies are still needed. 

In conclusion, the exploration of novel topological quantum materials with kagome lattice are now becoming intriguing and promising. In the future, more striking quantum properties are expected to be excavated and studied. It will undoubtedly promote the development of the emerging topological matters and the application in the fields of spintronics devices, quantum computation, and so on.

\textbf{BIOGRAPHICAL INFORMATION}

\textbf{Qi Wang} gained her Ph.D. degree from Renmin University of China in 2020. She is currently an assistant researcher at Shanghaitech laboratory for topological physics, Shanghaitech University. Her research interests mainly focus on the exploration of novel topological materials and the studies of physical properties under extreme conditions (ultra-low temperature, strong field, and high pressure), including the anomalous Hall effect, the topological Hall effect, and superconductivity. 

\textbf{Hechang Lei} obtained his Ph.D. degree from Institute of Solid State of Physics, Chinese Academy of Sciences in 2009. Now, he is a professor at the Department of Physics, Renmin University of China. His present research interests mainly focus on the exploration of exotic correlated topological matter and low-dimensional strongly correlated phenomena, such as superconductivity, frustrated magnetism, and other emergent low-dimensional quantum matter.

\textbf{Yanpeng Qi} obtained his Ph.D. degree from the Institute of Electrical Engineering, Chinese Academy of Sciences in 2011. Now, he is an assistant professor at the School of Physical Science and Technology, ShanghaiTech University. His current research interest focuses on the understanding of novel properties of quantum materials under high pressures, and the design and synthesis of new superconductors.

\textbf{Claudia Felser} is currently director at the Max Planck Institute for Chemical Physics of Solids in Dresden (Germany). Her research foci are the design and discovery of novel inorganic compounds, in particular, Heusler compounds for multiple applications and new topological quantum materials.

This work was supported by the National Key R\&D Program of China (Grants 2023YFA1607400, 2018YFA0704300, 2022YFA1403800, and 2023YFA1406500), the National Natural Science Foundation of China (Grants 52272265 and 12274459), the Shanghai Sailing Program (Grant 23YF1426800), and the Beijing Natural Science Foundation (Grant Z200005).

$\ast$ Correspondence should be addressed to Y. P. Qi (qiyp@shanghaitech.edu.cn) or H. C. Lei (hlei@ruc.edu.cn).

\clearpage
\begin{figure*}
	\centerline{\includegraphics[scale=0.5]{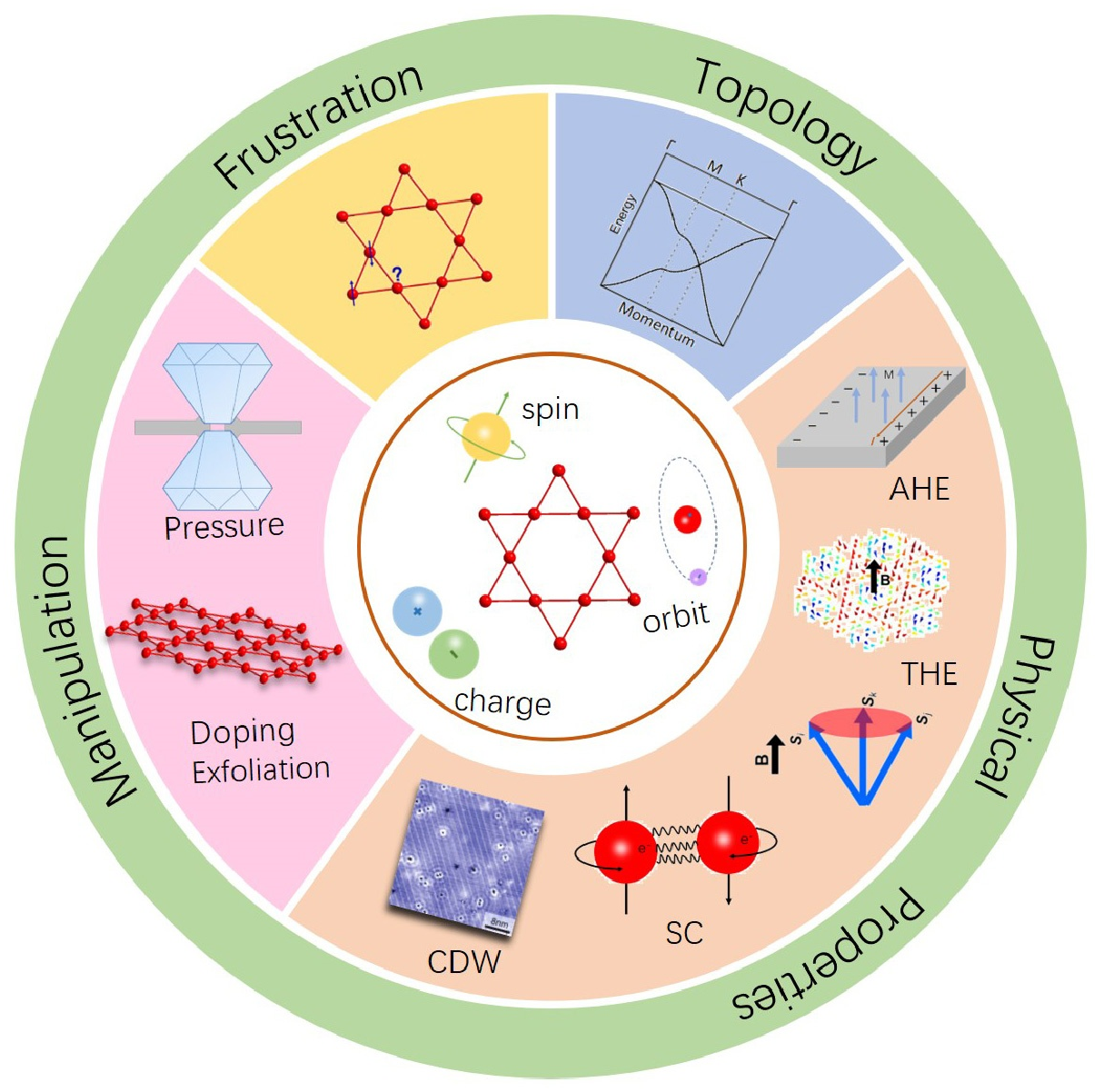}}
	\caption*{TOC}
\end{figure*}


\section{References}
\begin{thebibliography}{99}
	
\bibitem{1057} Qi, X.-L.; Zhang, S.-C. Topological insulators and superconductors. \textit{Rev. Mod. Phys.} \textbf{2011}, 83, 1057.  

\bibitem{021004} Bansil, A.; Lin, H.; Das, T. Colloquium: Topological band theory. \textit{Rev. Mod. Phys.} \textbf{2016}, 88, 021004.

\bibitem{240} Xie, Y.; Chen, L.; Chen, T.; Wang, Q.; Yin, Q.; Stewart, J. R.; Stone, M. B.; Daemen, L. L.; Feng, E.; Cao, H.; Lei, H.; Yin, Z.; MacDonald, A. H.; Dai, P. Spin excitations in metallic kagome lattice FeSn and CoSn. \textit{Commun Phys} \textbf{2021}, 4, 240.

\bibitem{287} Hall, E. H. On a new action of the magnet on electric currents. \textit{Am. J. Math.} \textbf{1879}, 2, 287.

\bibitem{301} Hall, E. H. On the new action of magnetism on a permanent electric current. \textit{Philos. Mag.} \textbf{1880}, 10, 301.

\bibitem{157} Hall, E. H. On the "Rotational coefficient" in nickel and cobalt. \textit{Philos. Mag.} \textbf{1881}, 12, 157.

\bibitem{1154} Karplus, R.; Luttinger, J. M. Hall effect in ferromagnetics. \textit{Phys. Rev.} \textbf{1954}, 95, 1154.

\bibitem{207208} Jungwirth, T.; Niu, Q.; MacDonald, A. H. Anomalous Hall Effect in Ferromagnetic Semiconductors. \textit{Phys. Rev. Lett.} \textbf{2002}, 88, 207208.

\bibitem{39} Smit, J. The spontaneous Hall effect in ferromagnetics II. \textit{Physica} \textbf{1958}, 24, 39.

\bibitem{4559} Berger, L. Side-jump mechanism for the Hall effect of ferromagnets. \textit{Phys. Rev. B} \textbf{1970}, 2, 4559.

\bibitem{156601} Ueda, K.; Iguchi, S.; Suzuki, T.; Ishiwata, S.; Taguchi, Y.; Tokura, Y. Topological Hall Effect in Pyrochlore Lattice with Varying Density of Spin Chirality. \textit{Phys. Rev. Lett.} \textbf{2012}, 108, 156601.

\bibitem{236802} Tang, E.; Mei, J. W.; Wen, X. G. High Temperature Fractional Quantum Hall States. \textit{Phys. Rev. Lett.} \textbf{2011}, 106, 236802.

\bibitem{115135} Wang, W.-S.; Li, Z.-Z.; Xiang, Y.-Y.; Wang, Q.-H. Competing electronic orders on kagome lattices at van Hove filling. \textit{Phys. Rev. B} \textbf{2013}, 87, 115135.

\bibitem{075135} Wang, Q.; Sun, S. S.; Zhang, X.; Pang, F.; Lei, H. C. Anomalous Hall effect in a ferromagnetic Fe$_{3}$Sn$_{2}$ single crystal with a geometrically frustrated Fe bilayer kagome lattice. \textit{Phys. Rev. B} \textbf{2016}, 94, 075135.

\bibitem{452202} Fenner, L. A.; Dee, A. A.; Wills, A. S. Non-collinearity and spin frustration in the itinerant kagome ferromagnet Fe$_{3}$Sn$_{2}$. \textit{J. Phys.: Condens. Matter} \textbf{2009}, 21, 452202.

\bibitem{638} Ye, L.; Kang, M.; Liu, J.; Cube, F. V.; Wicker, C. R.;  Suzuki, T.; Jozwiak, C.; Bostwick, A.; Bostwick, E.; Bell, D. C.; Fu, L.; Comin, R.; Checkelsky, J. G. Massive Dirac fermions in a ferromagnetic kagome metal. \textit{Nature} \textbf{2018}, 555, 638. 

\bibitem{186602} Neubauer, A.; Pfleiderer, C.; Binz, B.; Rosch, A.; Ritz, R.; Niklowitz, P. G.; Boni, P. Topological Hall Effect in the A Phase of MnSi. \textit{Phys. Rev. Lett.} \textbf{2009}, 102, 186602.

\bibitem{6456} Kurumaji, T.; Nakajima, T.; Hirschberger, M.; Kikkawa, A.; Yamasaki, Y.; Sagayama, H.; Nakao, H.; Taguchi, Y.; Arima, T.; Tokura, Y. Skyrmion lattice with a giant topological Hall effect in a frustrated triangular-lattice magnet. \textit{Science} \textbf{2019}, 365, 914.

\bibitem{1701144} Hou, Z.; Ren, W.; Ding, B.; Xu, G.; Wang, Y.; Yang, B.; Zhang, Q.; Zhang, Y.; Liu, E.; Xu, F.; Wang, W.; Wu, G.; Zhang, X.; Shen, B.; Zhang, Z. Observation of Various and Spontaneous Magnetic Skyrmionic Bubbles at Room Temperature in a Frustrated Kagome Magnet with Uniaxial Magnetic Anisotropy. \textit{Adv. Mater.} \textbf{2017}, 29, 1701144.

\bibitem{017101} Wang, Q.; Yin, Q. W.; Lei, H. C. Giant topological Hall effect of ferromagnetic kagome metal Fe$_{3}$Sn$_{2}$. \textit{Chin. Phys. B} \textbf{2020}, 29, 017101.

\bibitem{196604} Li, Y. M.; Wang, Q.; Schmitt, L. D.; Guguchia, Z.; Desautels, R. D.; Yin, J. X.; Du, Q.; Ren, W.; Zhao, X.; Zhang, Z.; Zaliznyak, I. A.; Petrovic, C.; Yin, W.; Hasan, M. Z.; Lei, H. C.; Tranquada, J. M. Magnetic-field Control of Topological Electronic Response near Room Temperature in Correlated Kagome Magnets. \textit{Phys. Rev. Lett.} \textbf{2019}, 123, 196604. 

\bibitem{096401} Lin, Z.; Choi, J.-H.; Zhang, Q.; Qin, W.; Yi, S.; Wang, P.; Li, L.; Wang, Y.; Zhang, H.; Sun, Z.; Wei, L.; Zhang, S.; Guo, T.; Lu, Q.; Cho, J.-H.; Zeng, C.; Zhang, Z. Flatbands and Emergent Ferromagnetic Ordering in Fe$_{3}$Sn$_{2}$ Kagome Lattices. \textit{Phys. Rev. Lett.} \textbf{2018}, 121, 096401.

\bibitem{205101} Wan, X.; Turner, A. M.; Vishwanath, A.; Savrasov, S. Y. Topological semimetal and Fermi arc surface states in the electronic structure of pyrochlore iridates. \textit{Phys. Rev. B} \textbf{2011}, 83, 205101.

\bibitem{186806} Xu, G.; Weng, H.; Wang, Z.; Dai, X.; Fang, Z. Chern semimetal and the quantized anomalous Hall effect in HgCr$_{2}$Se$_{4}$. \textit{Phys. Rev. Lett.} \textbf{2011}, 107, 186806.

\bibitem{3681} Wang, Q.; Xu, Y. F.; Lou, R.; Liu, Z.; Li, M.; Huang, Y.; Shen, D.; Weng, H. M.; Wang, S. C.; Lei, H. C. Large intrinsic anomalous Hall effect in halfmetallic ferromagnet Co$_{3}$Sn$_{2}$S$_{2}$ with magnetic Weyl fermions. \textit{Nat. Commun.} \textbf{2018}, 9, 3681.

\bibitem{1125} Liu, E.; Sun, Y.; Kumar, N.; Muechler, L.; Sun, A.; Jiao, L.; Yang, S. Y.; Liu, D.; Liang, A.; Xu, Q.; Kroder, J.; S\"{u}{\ss}, V.; Borrmann, H.; Shekhar, C.; Wang, Z.; Xi, C.; Wang, W.; Schnelle, W.; Wirth, S.; Chen, Y.; Goennenwein, S. T. B.; Felser, C. Giant anomalous Hall effect in a ferromagnetic kagome-lattice semimetal. \textit{Nat. Phys.} \textbf{2018}, 14, 1125.

\bibitem{1282} Liu, D. F.; Liang, A. J.; Liu, E. K.; Xu, Q. N.; Li, Y. W.; Chen, C.; Pei, D.; Shi, W. J. ; Mo, S. K.; Dudin, P.; Kim, T.; Cacho, C.; Li, G.; Sun, Y.; Yang, X. L.; Liu, Z. K.; Parkin, S. S. P.; Felser, C.; Chen, Y. L. Magnetic Weyl semimetal phase in a Kagome crystal. \textit{Science} \textbf{2019}, 365, 1282.

\bibitem{1278} Morali, N.; Batabyal, R.; Nag, P. K.; Liu, E.; Xu, Q.; Sun, Y.; Yan, B.; Felser, C.; Avraham, N.; Beidenkopf, H. Fermi-arc diversity on surface terminations of the magnetic Weyl semimetal Co$_{3}$Sn$_{2}$S$_{2}$. \textit{Science} \textbf{2019}, 365, 1286.

\bibitem{443} Yin, J. X.; Zhang, S. S.; Chang, G. Q.; Wang, Q.; Tsirkin, S. S.; Guguchia, Z.; Lian, B.; Zhou, H.; Jiang, K.; Belopolski, I.; Shumiya, N.; Multer, D.; Litskevich, M.; Cochran, T. A.; Lin, H.; Wang, Z.; Neupert, T.; Jia, S.; Lei, H. C.; Hasan, M. Z. Negative flat band magnetism in a spin-orbit-coupled correlated kagome magnet. \textit{Nat. Phys.} \textbf{2019}, 15, 443.

\bibitem{014416} Wang, Q.; Neubauer, K. J.; Duan, C.; Yin, Q. W.; Fujitsu, S.; Hosono, H.; Ye, F.; Zhang, R.; Chi, S.; Krycka, K.; Lei, H. C.; Dai, P. C. Field-induced topological Hall effect and double-fan spin structure with a $c$-axis component in the metallic kagome antiferromagnetic compound YMn$_{6}$Sn$_{6}$. \textit{Phys. Rev. B} \textbf{2021}, 103, 014416.

\bibitem{246602} Ma, W.; Xu, X.; Yin, J.-X.; Yang, H.; Zhou, H.; Cheng, Z.-J.; Huang, Y.; Qu, Z.; Wang, F.; Hasan, M. Z.; Jia, S. Rare Earth Engineering in RMn$_{6}$Sn$_{6}$ (R = Gd-Tm, Lu). \textit{Phys. Rev. Lett.} \textbf{2021}, 126, 246602. 

\bibitem{533} Yin, J.-X.; Ma, W.; Cochran, T. A.; Xu, X.; Zhang, S. S.; Tien, H.-J.; Shumiya, N.; Cheng, G.; Jiang, K.; Lian, B.; Song, Z.; Chang, G.; Belopolski, I.; Multer, D.; Litskevich, M.; Cheng, Z.-J.; Yang, X. P.; Swidler, B.; Zhou, H.; Lin, H.; Neupert, T.; Wang, Z.; Yao, N.; Chang, T.-R.; Jia, S.; Hasan, M. Z. Quantum-limit Chern topological magnetism in TbMn$_{6}$Sn$_{6}$. \textit{Nature} \textbf{2020}, 583, 533.

\bibitem{3129} Li, M.; Wang, Q.; Wang, G.; Yuan, Z.; Song, W.; Lou, R.; Liu, Z.; Huang, Y.; Liu, Z.; Lei, H. C.; Yin, Z. P.; Wang, S. C. Dirac cone, flat band and saddle point in kagome magnet YMn$_{6}$Sn$_{6}$. \textit{Nat. Commun.} \textbf{2021}, 12, 3129.

\bibitem{212} Nakatsuji, S.; Kiyohara, N.; Higo, T. Large anomalous Hall effect in a non-collinear antiferromagnet at room temperature. \textit{Nature} \textbf{2015}, 527, 212.

\bibitem{1090} Kuroda, K.; Tomita, T.; Suzuki, M.-T.; Bareille, C.; Nugroho, A. A.; Goswami, P.; Ochi, M.; Ikhlas, M.; Nakayama, M.; Akebi, S.; Noguchi, R.; Ishii, R.; Inami, N.;  Ono, K.; Kumigashira, H.; Varykhalov, A.; Muro, T.; Koretsune, T.; Arita, R.; Shin, S.; Kondo, T.; Nakatsuji, S. Evidence for magnetic Weyl fermions in a correlated metal. \textit{Nat. Mater.} \textbf{2017}, 16, 1090. 

\bibitem{163} Kang, M.; Ye, L.; Fang, S.; You, J. S.; Levitan, A.; Han, M.; Facio, J. I.; Jozwiak, C.; Bostwick, A.; Rotenberg, E.; Chan, M. K.; McDonald, R. D.; Graf, D.;  Kaznatcheev, K.; Vescovo, E.; Bell, D. C.; Kaxiras, E.; van den Brink, J.; Richter, M.; Ghimire, M. P.; Checkelsky, J. G.; Comin, R. Dirac fermions and flat bands in the ideal kagome metal FeSn. \textit{Nat. Mater.} \textbf{2020}, 19, 163.

\bibitem{4002} Liu, Z.; Li, M.; Wang, Q.; Wang, G.; Wen, C.; Jiang, K.; Lu, X.; Yan, S.; Huang, Y.; Shen, D.; Yin, J.-X.; Wang, Z.; Yin, Z.; Lei, H. C.; Wang, S. C. Orbital-selective Dirac fermions and extremely flat bands in frustrated kagome-lattice metal CoSn. \textit{Nat. Commun.} \textbf{2020}, 11, 4002.

\bibitem{4003} Yin, J. X.; Shumiya, N.; Mardanya, S.; Wang, Q.; Zhang, S. S.; Tien, H.-J.; Multer, D.; Jiang, Y.; Cheng, G.; Yao, N.; Wu, S.; Wu, D.; Deng, L.; Ye, Z.; He, R.; Chang, G.; Liu, Z.; Jiang, K.; Wang, Z.; Neupert, T.; Agarwal, A.; Chang, T.-R.; Chu, C.-W.; Lei, H. C.; Hasan, M. Z. Fermion-boson many-body interplay in a frustrated kagome paramagnet. \textit{Nat. Commun.} \textbf{2020}, 11, 4003. 

\bibitem{094407} Ortiz, B. R.; Gomes, L. C.; Morey, J. R.; Winiarski, M.; Bordelon, M.; Mangum, J. S.; Oswald, I. W. H.; Rodriguez-Rivera, J. A.; Neilson, J. R.; Wilson, S. D.;  Ertekin, E.; McQueen, T. M.; Toberer, E. S. New kagome prototype materials: discovery of KV$_{3}$Sb$_{5}$, RbV$_{3}$Sb$_{5}$, and CsV$_{3}$Sb$_{5}$. \textit{Phys. Rev. Mater.} \textbf{2019}, 3, 094407.

\bibitem{247002} Ortiz, B. R.; Teicher, S. M. L.; Hu, Y.; Zuo, J. L.; Sarte, P. M.; Schueller, E. C.; Abeykoon, A. M. M.; Krogstad, M. J.; Rosenkranz, S.; Osborn, R.; Seshadri, R.;  Balents, L.; He, J.; Wilson, S. D. CsV$_{3}$Sb$_{5}$: A $Z_{2}$ Topological Kagome Metal with a Superconducting Ground State. \textit{Phys. Rev. Lett.} \textbf{2020}, 125, 247002.

\bibitem{037403} Yin, Q. W.; Tu, Z. J.; Gong, C. S.; Fu, Y.; Yan, S. H.; Lei, H. C. Superconductivity and Normal-State Properties of Kagome Metal RbV$_{3}$Sb$_{5}$ Single Crystals. \textit{Chin. Phys. Lett.} \textbf{2021}, 38, 037403.

\bibitem{2102813} Wang, Q.; Kong, P.; Shi, W.; Pei, C.; Wen, C.; Gao, L.; Zhao, Y.; Yin, Q.; Wu, Y.; Li, G.; Lei, H. C.; Li, J.; Chen, Y. L.; Yan, S. C.; Qi, Y. P. Charge Density Wave Orders and Enhanced Superconductivity under Pressure in the Kagome Metal CsV$_{3}$Sb$_{5}$. \textit{Adv. Mater.} \textbf{2021}, 33, 2102813.

\bibitem{031026} Liang, Z.; Hou, X.; Zhang, F.; Ma, W.; Wu, P.; Zhang, Z.; Yu, F.; Ying, J.-J.; Jiang, K.; Shan, L.; Wang, Z.; Chen, X.-H. Three-Dimensional Charge Density Wave and Surface-Dependent Vortex-Core States in a Kagome Superconductor CsV$_{3}$Sb$_{5}$. \textit{Phys. Rev. X} \textbf{2021}, 11, 031026.

\bibitem{031050} Li, H.; Zhang, T. T.; Yilmaz, T.; Pai, Y. Y.; Marvinney, C. E.; Said, A.; Yin, Q. W.; Gong, C. S.; Tu, Z. J.; Vescovo, E.; Nelson, C. S.; Moore, R. G.; Murakami, S.; Lei, H. C.; Lee, H. N.; Lawrie, B. J.; Miao, H. Observation of Unconventional Charge Density Wave without Acoustic Phonon Anomaly in Kagome Superconductors AV$_{3}$Sb$_{5}$ (A = Rb, Cs). \textit{Phys. Rev. X} \textbf{2021}, 11, 031050.

\bibitem{247462} Song, D.; Zheng, L.; Yu, F.; Li, J.; Nie, L.; Shan, M.; Zhao, D.; Li, S. J.; Kang, B. L.; Wu, Z. M.; Zhou, Y. B.; Sun, K. L.; Liu, K.; Luo, X. G.; Wang, Z. Y.; Ying, J. J.; Wan, X. G.; Wu, T.; Chen, X. H. Orbital ordering and fluctuations in a kagome superconductor CsV$_{3}$Sb$_{5}$. \textit{Sci. China Phys. Mech. Astron.} \textbf{2022}, 65, 247462.

\bibitem{490} Teng, X.; Chen, L.; Ye, F.; Rosenberg, E.; Liu, Z.; Yin, J.-X.; Jiang, Y.-X.; Oh, J. S.; Hasan, M. Z.; Neubauer, K. J.; Gao, B.; Xie, Y.; Hashimoto, M.; Lu, D.;  Jozwiak, C.; Bostwick, A.; Rotenberg, E.; Birgeneau, R. J.; Chu, J.-H.; Yi, M.; Dai, P. Discovery of charge density wave in a kagome lattice antiferromagnet. \textit{Nature} \textbf{2022}, 609, 490.

\bibitem{216402} Arachchige, H. W. S.; Meier, W. R.; Marshall, M.; Matsuoka, T.; Xue, R.; McGuire, M. A.; Hermann, R. P.; Cao, H.; Mandrus, D. Charge Density Wave in Kagome Lattice Intermetallic ScV$_{6}$Sn$_{6}$. \textit{Phys. Rev. Lett.} \textbf{2022}, 129, 216402.

\bibitem{105} Wu, Y.; Wang, Q.; Zhou, X.; Wang, J.; Dong, P.; He, J.; Ding, Y.; Teng, B.; Zhang, Y.; Li, Y.; Zhao, C.; Zhang, H.; Liu, J.; Qi, Y. P.; Watanabe, K.; Taniguchi, T.; Li, J. Nonreciprocal Charge Transport in Topological Kagome Superconductor CsV$_{3}$Sb$_{5}$. \textit{npj Quantum Mater.} \textbf{2022}, 7, 105. 

\bibitem{eabb6003} Yang, S.-Y.; Wang, Y.; Ortiz, B. R.; Liu, D.; Gayles, J.; Derunova, E.; Gonzalez-Hernandez, R.; Smejkal, L.; Chen, Y.; Parkin, S. S. P.; Wilson, S. D.; Toberer, E. S.; McQueen, T.; Ali, M. N. Giant, unconventional anomalous Hall effect in the metallic frustrated magnet candidate, KV$_{3}$Sb$_{5}$. \textit{Sci. Adv.} \textbf{2020}, 6, eabb6003.

\bibitem{1038} Jiang, Y.-X.; Yin, J.-X.; Denner, M. M.; Shumiya, N.; Ortiz, B. R.; Xu, G.; Guguchia, Z.; He, J.; Hossain, M. S.; Liu, X.; Ruff, J.; Kautzsch, L.; Zhang, S. S.; Chang, G.; Belopolski, I.; Zhang, Q.; Cochran, T. A.; Multer, D.; Litskevich, M.; Cheng, Z.-J.; Yang, X. P.; Wang, Z.; Thomale, R.; Neupert, T.; Wilson, S. D.; Hasan, M. Z. Unconventional chiral charge order in kagome superconductor KV$_{3}$Sb$_{5}$.\textit{Nat. Mater.} \textbf{2021}, 20, 1353.

\bibitem{222} Chen, H.; Yang, H.; Hu, B.; Zhao, Z.; Yuan, J.; Xing, Y.; Qian, G.; Huang, Z.; Li, G.; Ye, Y.; Ma, S.; Ni, S.; Zhang, H.; Yin, Q.; Gong, C.; Tu, Z.; Lei, H.; Tan, H.; Zhou, S.; Shen, C.; Dong, X.; Yan, B.; Wang, Z.; Gao, H.-J. Roton pair density wave in a strong-coupling kagome superconductor. \textit{Nature} \textbf{2021}, 599, 222. 

\bibitem{59} Nie, L.; Sun, K.; Ma, W.; Song, D.; Zheng, L.; Liang, Z.; Wu, P.; Yu, F.; Li, J.; Shan, M.; Zhao, D.; Li, S.; Kang, B.; Wu, Z.; Zhou, Y.; Liu, K.; Xiang, Z.; Ying, J.; Wang, Z.; Wu, T.; Chen, X. H. Charge-density-wave-driven electronic nematicity in a kagome superconductor. \textit{Nature} \textbf{2022}, 604, 59.

\bibitem{207002} Fu, Y.; Zhao, N.; Chen, Z.; Yin, Q.; Tu, Z.; Gong, C.; Xi, C.; Zhu, X.; Sun, Y.; Liu, K.; Lei, H. Quantum Transport Evidence of Topological Band Structures of Kagome Superconductor CsV$_{3}$Sb$_{5}$. \textit{Phys. Rev. Lett.} \textbf{2021}, 127, 207002.

\bibitem{6727} Xiang, Y.; Li, Q.; Li, Y.; Xie, W.; Yang, H.; Wang, Z.; Yao, Y.; Wen, H.-H. Twofold symmetry of $c$-axis resistivity in topological kagome superconductor CsV$_{3}$Sb$_{5}$ with in-plane rotating magnetic field. \textit{Nat. Commun.} \textbf{2021}, 12, 6727.

\bibitem{043001} Wang, Q.; Qiu, X.-L.; Pei, C.; Gong, B.-C.; Gao, L.; Zhao, Y.; Cao, W.; Li, C.; Zhu, S.; Zhang, M.; Chen, Y. L.; Liu, K.; Qi, Y. P. Superconductivity emerging from a pressurized van der Waals kagome material Pd$_{3}$P$_{2}$S$_{8}$, \textit{New J. Phys.} \textbf{2023}, 25, 043001.

\bibitem{20998} Park, S.; Kang, S.; Kim, H.; Lee, K. H.; Kim, P.; Sim, S.; Lee, N.; Karuppannan, B.; Kim, J.; Kim, J.; Sim, K. I.; Coak, M. J.; Noda, Y.; Park, C.-H.; Kim, J. H.; Park, J.-G. Kagome van-der-Waals Pd$_{3}$P$_{2}$S$_{8}$ with flat band. \textit{Sci. Rep.} \textbf{2020}, 10, 20998.

\bibitem{155115} Yan, S.; Gong, B.-C.; Wang, L.; Wu, J.; Yin, Q.; Cao, X.; Zhang, X.; Liu, X.; Lu, Z.-Y.; Liu, K.; Lei, H. C. Evolution of ultraflat band in the van der Waals kagome semiconductor Pd$_{3}$P$_{2}$(S$_{1-x}$Se$_{x}$)$_{8}$. \textit{Phys. Rev. B} \textbf{2022}, 105, 155115.

\bibitem{115106} Muechler, L.; Liu, E.; Gayles, J.; Xu, Q.; Felser, C.; Sun, Y. Emerging chiral edge states from the confinement of a magnetic Weyl semimetal in Co$_{3}$Sn$_{2}$S$_{2}$. \textit{Phys. Rev. B} \textbf{2020}, 101, 115106.

\bibitem{014410} Zhang, Z.; You, J.-Y.; Ma, X.-Y.; Gu, B.; Su, G. Kagome quantum anomalous Hall effect with high Chern number and large band gap. \textit{Phys. Rev. B} \textbf{2021}, 103, 014410.


\end{thebibliography}
\end{document}